\documentstyle[12pt,epsfig]{article}

\newcommand{\ra}{\right >}
\newcommand{\la}{\left <}

\begin{document}

\title{Stick-slip Transition in the Scalar Arching Model}

\author{P. Claudin and J.-P. Bouchaud\\
Service de Physique de l'Etat Condens\'e,\\
CEA, Orme des Merisiers,\\
91191 Gif-sur-Yvette, Cedex France.\\
}

\date{\today}
\setcounter{page}{0}
\maketitle

\vspace{1cm}

\begin{abstract}
When some granular material contained into a silo is pushed upwards with a piston, an
irregular stick-slip
motion of the system of grains is observed. We show how one can adapt
the `Scalar Arching Model' ({\sc sam}) -- proposed as a  model for giant stress
fluctuations in silos -- in order to describe this stick-slip phenomenon.
As a function of the sensitivity of the system to mechanical noise, the system
exhibit two different phases: a `jammed' phase, and a `sliding' phase where
irregular stick-slip is observed. We analyze the transition, which is found to be
of mean-field type, and study the statistical properties of the intermittent
stick-slip motion. \end{abstract}

\thispagestyle{empty}
\vspace{8cm}
\newpage

Stick-slip motion is a very common phenomenon which occurs when two solids slide on
each other, and has been much studied in the recent years in connection with solid
friction \cite{Baumberger,Ciliberto}. Stick-slip motion also occurs in granular
materials \cite{Gollub,Jussieu,Touria}. For example, the Jussieu group has
performed the following experiment \cite{Jussieu,Touria}: a 2D vertical cell
containing aluminium beads is pushed upwards by a piston through to a spring tightened
at a constant speed. A very irregular stick-slip motion is observed, which can be
attributed to the fact that force propagation in granular media is strongly
inhomogeneous: most of the force is concentrated on stress paths (or arches). Some
configurations of these paths propagate all the extenal pushing force $F$ to the
walls, and jamming occurs. Other configurations propagate the pushing force right
up to the free surface, and the beads move upwards (slip). The temporal
fluctuations of these stress paths then lead to an intermittent stick-slip motion,
which we attempt to model here using the `Scalar Arching Model' ({\sc sam}). This
model was proposed to model giant stress fluctuations in silos \cite{CB}. The {\sc
sam} is an extension of the Chicago group's stochastic model \cite{Liu,Copper}
which allows for the formation of arches. We have shown how small perturbations could
lead to sudden transformations of these arches, and thus to large fluctuations of the
weight on the bottom plate of a silo. The idea is to adapt the same model to the
experimental situation described above, and to quantify how the formation of
arches can indeed lead to very irregular stick-slip motions.

\begin{figure}[hbt]
\begin{center}
\epsfysize=6cm
\epsfbox{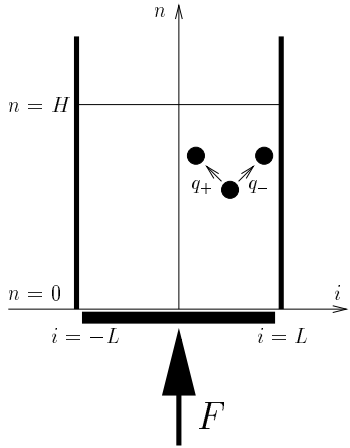}
\caption{\small Beads are confined into a 2D silo.
An upwards force $F$ is applied on a piston at the bottom of the silo.
Such a system of beads gives rise to an irregular stick-slip motion.
In the {\sc sam} model, beads propagate forces according to rules which combines
random propagation (encoded by the random numbers $q_\pm$) and arch formation. In all
the present paper, simulations have been performed with $L=30$ (bead radii), which is
close to the experimental system of \protect\cite{Jussieu,Touria}. \label{shema}}
\end{center}
\end{figure}

Let us briefly recall the main features of the {\sc sam}. The granular packing is
represented in 2D by a regular lattice: each site is a `grain' labelled by two
integers $(i,n)$ giving its horizontal and vertical coordinate. We neglect the weight
of the grains compared with the applied force $F$, and focus on the transmission of
$F$ through the grains from the bottom of the cell ($n=0$) to the walls ($i=\pm L$)
and the free surface ($n=H$), see figure \ref{shema}.
Each grain then supports the force $w$ of its two downstairs neighbours, and shares
its own load randomly between its two upstairs neighbours.
The corresponding scalar equation for force propagation is thus: 
\begin{equation} 
w(i,n) = q_+(i+1,n-1)w(i+1,n-1) + q_-(i-1,n-1)w(i-1,n-1)
\end{equation}
$q_\pm(i,n)$ is the fraction of the force transmitted to the the grain $(i\mp 1,n+1)$,
and is a random variable between $0$ and $1$, subject to the conservation constraint
$q_+(i,n)+q_-(i,n)=1$. These random coefficients model the randomness of the local packing,
size and shape of the grains, etc. At this stage, the model is the one considered in
\cite{Liu}, \cite{Copper}, albeit upside down. We now include a `local slip
condition': when the shear on a given grain is too strong, the grain can slip and
lose its contact with its neighbours opposite to the direction of the shear. More
precisely, we introduced a threshold $R_c$ such that \begin{eqnarray}
q_+(i,n)=1-q_-(i,n)=0 & \qquad \mbox{\ if \ } & {w_--w_+ \over w(i,n)} \ge R_c \\
q_-(i,n)=1-q_+(i,n)=0 & \qquad \mbox{\ if \ } & {w_+-w_- \over w(i,n)} \ge R_c
\end{eqnarray}
where $w_\pm=q_\pm(i\pm1,n-1)w(i\pm1,n-1)$. These rules lead to arch formation
\cite{CB}.

The walls play a crucial
r\^ole in the stick-slip process, since the friction forces there can balance
the external force $F$ and allow the system to jam.  We thus introduced a new
parameter, the `jamming ability' $\alpha$, such that with probability $\alpha$ the
load $w(\pm L,n)$ is completely `absorbed' by the wall which balances all the
force carried by the grain. With probability $1-\alpha$, we apply the same rule as
in the bulk, i.e. for $i=\pm L$, the fraction $q_\mp(\pm L,n)$ of the load $w(\pm
L,n)$ hits the wall. In addition to the random $q_\pm(i,n)$, we thus define $2(H+1)$
uniform random numbers $\alpha_\pm (n)$ which, compared to $\alpha$, decide whether the site $(i=\pm L,n)$ is `absorbing' (i.e. if the grain is supported only by the wall) or not. 
In the following, the $q_\pm(i,n)$'s are
also chosen uniformly between $0$ and $1$, although other choices lead to the same
qualitative conclusions.

In order to reproduce the overall stick-slip motion of the assembly of grains, we
propose to capture the different physical phenomena which occur as follows. For a
given applied external force $F$, and a given set of random numbers $q_\pm(i,n)$ and
$\alpha_\pm (n)$,\\ 
$\bullet$ we calculate the total
forces on the walls $F_w$ and on the free surface of the silo $F_{fs}$. Obviously,
$F=F_w+F_{fs}$.\\ 
$\bullet$ if $F_{fs}=0$, the grains do not move, corresponding to a {\it stick}
situation. We then increase the applied force $F$ by some fixed amount $\Delta F$ and
the time $t$ by $\Delta t$. In order to mimic the mechanical noise which
necessarily occurs when the external load is increased and might trigger some local 
rearrangements, we also change a fraction $p$ of all the random numbers and
recalculate $F_{fs}$.\\ 
$\bullet$ if $F_{fs} > 0$, the equilibrium condition for the top grains is not
satisfied, which means that grains are moving. It is a {\it slip} situation.
Correspondingly, the spring loosens and the applied force $F$ is decreased by
$\Delta F$. We also change {\it all} random numbers (because the flow motion
completely rearranges the packing). The flow stops when a randomly chosen
configuration has enough `anchoring' sites at the walls to yield $F_{fs}=0$.

\begin{figure}[p]
\begin{center}
\epsfysize=6cm
\epsfbox{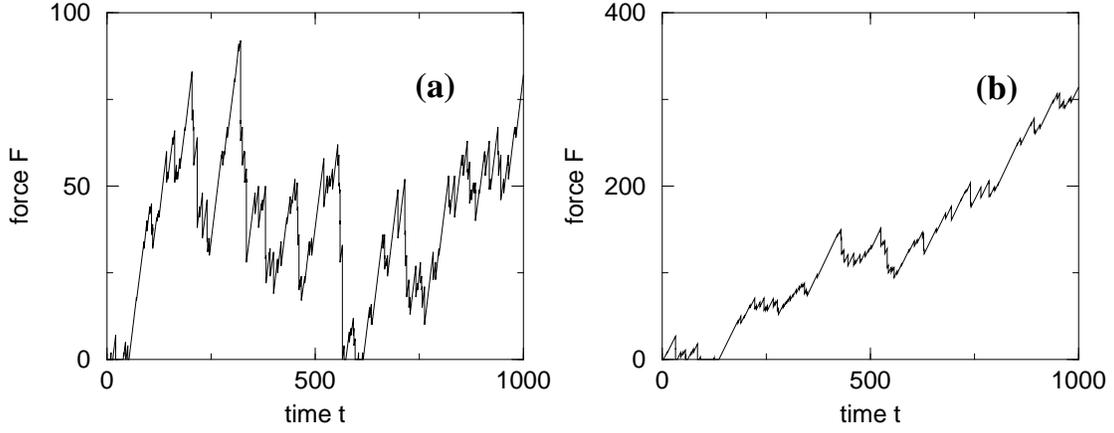}
\caption{\small These plots show the temporal evolution of the applied force $F$ for
(a) $\alpha = 0.83 < \alpha_c$ (sliding phase) and
(b) $\alpha = 0.85 > \alpha_c$ (jammed phase).
Both plots have been obtained with $R_c=0.5$, $p=0.01$
and $b=1$ for wich $\alpha_c \sim 0.838$.
\label{phases}}
\end{center}
\end{figure}

The simulation starts at $t=0$ with $F=0$ and lets $F$ increase progressively.
It is important to note that our model is actually purely static: no dynamics is
explicitly included. Therefore, the motion of the grains during a slipping event is
assumed to be infinitely quick on the scale of sticking events ($t$ is thus  kept
constant during slipping events). We thus actually describe only sticking situations,
separated by slipping events which have two effects: untighten the spring governing
the external force $F$, and reinitialize the structure of the packing (i.e. the
random numbers). As seen on figure \ref{phases}, such an `algorithm' indeed  leads 
to an irregular stick-slip motion. Note that for $\alpha=0$ the probability that $F_{fs}$
vanishes is exponentially small in $H$. Physically, this means that, in order to
resist to the external force $F$, the system of beads must generate arches which are
strongly `anchored' by the walls.

\begin{figure}[p]
\begin{center}
\epsfysize=6cm
\epsfbox{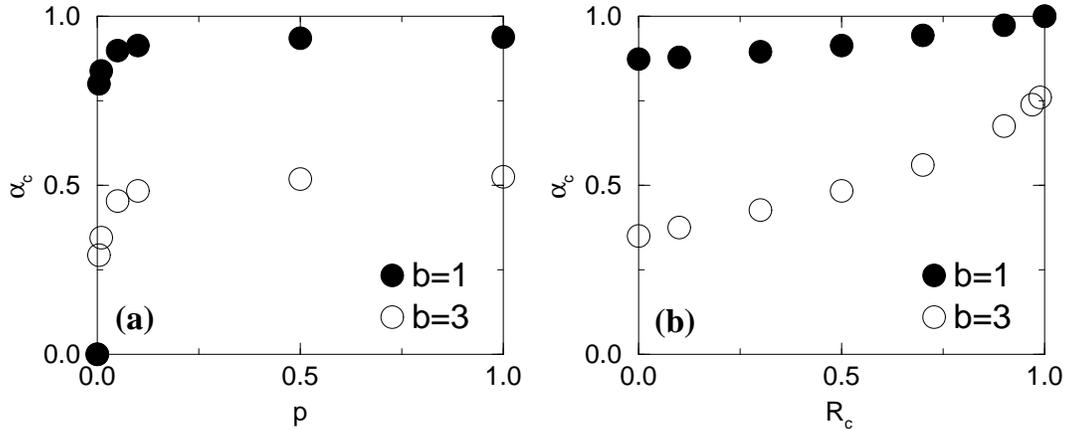}
\caption{\small The curve $\alpha = \alpha_c (p)$ separates the sliding phase (below)
from the jammed phase (above). The phase diagram (a) has been plotted for
$R_c=0.5$ and two aspect ratios, $b=1$ and $b=3$. Alternatively,
we can represent the phase diagram by the curve $\alpha = \alpha_c (R_c)$.
The parameters chosen for the figure (b) are $p=0.1$ and again, $b=1$ and $b=3$.
\label{phasediag}}
\end{center}
\end{figure}

The model is controlled by four parameters. The first two -- namely $R_c$, the
threshold of the {\sc sam}, and $b$ the aspect ratio of the cell -- are of secondary
importance: they do not affect the general features of our results. On the other
hand, the jamming ability $\alpha$ of the walls and the r\^ole of the mechanical
noise which modifies the local structure of the packing, measured by $p$. For
$R_c$ and $b$ fixed, depending on the values of $\alpha$ and $p$, two distinct
phases are found.  For small $p$ or large $\alpha$, the system is jammed in the sense
that although the grains move from time to time, $F$ goes to infinity as time
increases. This phase can be described by the average rate of increase of $F$,
$s=F(t)/t$. On the other hand, for large $p$ or small $\alpha$, the system slips 
very easily. This sliding phase is
characterized by the fact that $F(t)$ always goes back to $0$. The
relevant quantity in this phase is the delay $\tau$ between
two consecutive times where $F$ vanishes.  Figure \ref{phases} shows some typical
plots of $F(t)$ in these two phases. The transition between the two phases occurs
for a critical value of $\alpha=\alpha_c(p)$, which allows us to obtain the phase
boundaries, as plotted on figure \ref{phasediag}. Experimentally, these two regimes
could be reached by preparing the system in different ways. A compact system would
provide a rigid structure (small $p$). By contrast, a loose packing would be subject
to large rearrangements (large $p$). Similarly, the state of the walls allows one to
change the value of $\alpha$. An experimental recordings of $F(t)$ actually show parts
of the two different regimes \cite{Jussieu,Touria}.


\begin{figure}[p]
\begin{center}
\epsfysize=7cm
\epsfbox{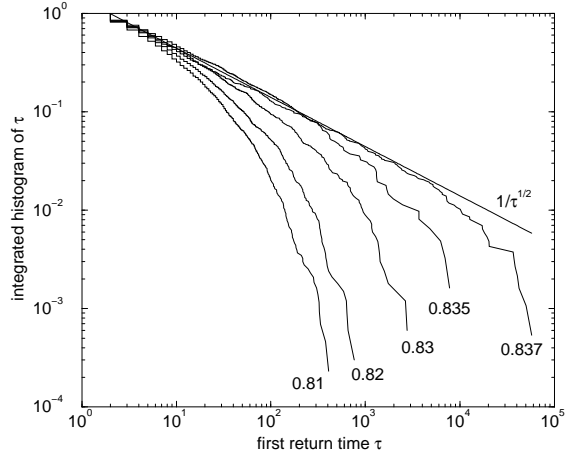}
\caption{\small These curves represent integrated histograms of the first return time $\tau$, 
i.e. the interval of time between two times where $F$ vanishes. They have been computed
with $R_c=0.5$, $p=0.01$, $b=1$, and with differents values
of $\alpha$ indicated on the plot. As $\alpha \to \alpha_c$, this histogram gets
broader and broader, and tends to the power law $\tau^{-1/2}$ which is characteristic
of the return time of simple random walks. \label{histotau}} \end{center}
\end{figure}

\begin{figure}[p]
\begin{center}
\epsfysize=7cm
\epsfbox{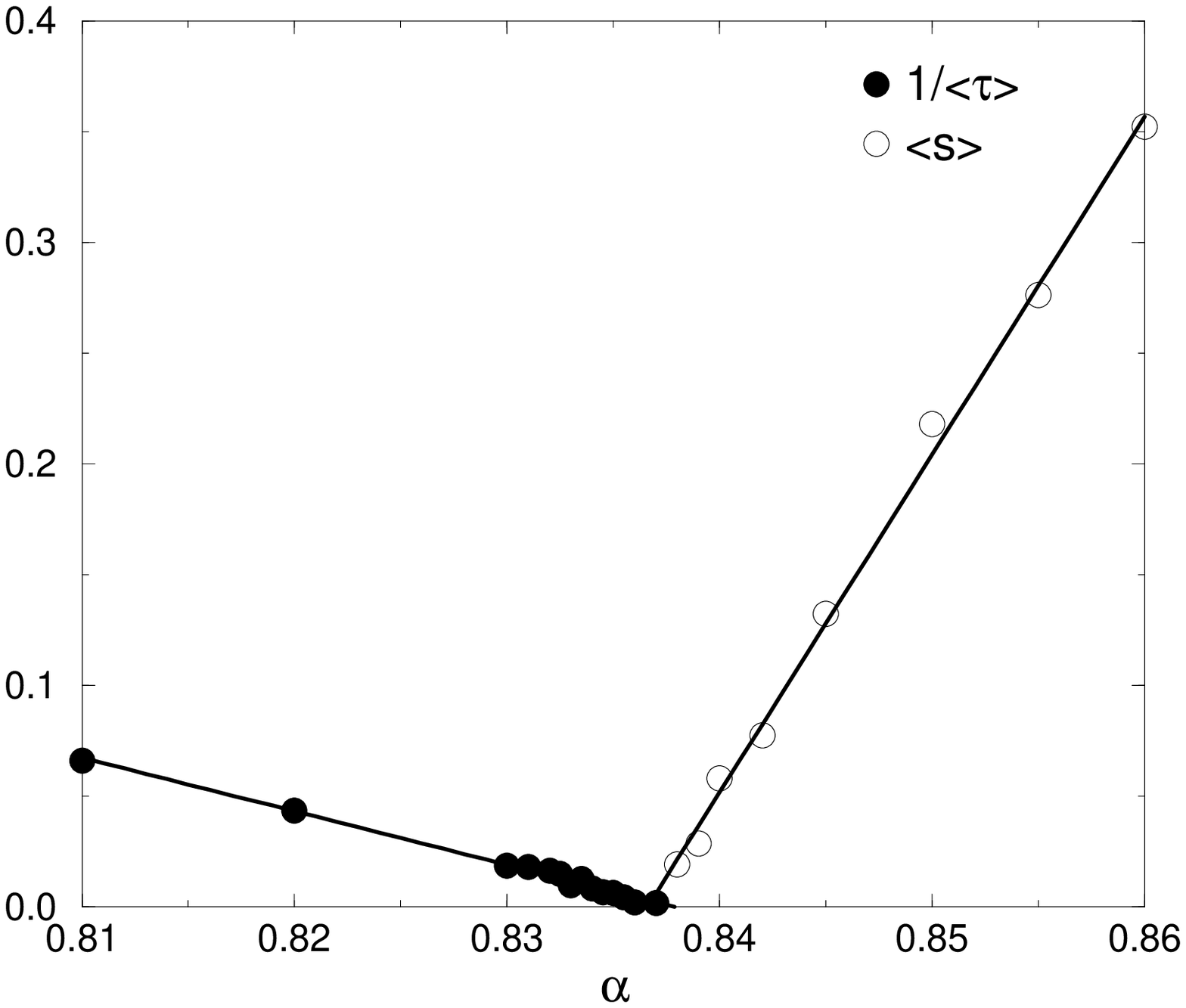}
\caption{\small Below the transition, the system is characterized by the averaged
first return time $\la \tau \ra$, and above it by the averaged slope $\la s \ra$ of
the applied force $F$ versus time. Near the transition,
we find that $\la \tau \ra$ diverges like $1/(\alpha_c-\alpha)$ and that $\la s \ra$ grows like
$\alpha-\alpha_c$. This plot has been computed with $R_c=0.5$, $p=0.01$ and $b=1$.
Linear regressions for $1/ \la \tau \ra$ and $\la s \ra$ give respectively
$\alpha_c =0.839 \pm 0.004$ and $\alpha_c =0.837 \pm 0.004$.
\label{trans}}
\end{center}
\end{figure}

We studied how the system behaves near criticality. On figure \ref{histotau}, we show
the integrated histogram of $\tau$ for $R_c$, $b$ and $p$ fixed, and for
different values of $\alpha$. We see that $\tau$ tends to be power-law distributed as
$\alpha \to \alpha_c$, with an exponent $-1/2$. Note that, as argued below, this
power-law corresponds to the first return probability of a one dimensional random
walk. $\la \tau \ra$ diverges for $\alpha=\alpha_c$; we found numerically that
$\la \tau \ra \propto 1/(\alpha_c-\alpha)$ for $\alpha < \alpha_c$. In the same way,
we found that the average slope of $F$ versus time behaves like $\la s \ra \propto
(\alpha-\alpha_c)$ for $\alpha > \alpha_c$ (see figure \ref{trans}).


These critical laws can be understood within a simple mean field analysis. Neglecting
the correlations, the temporal evolution of $F$ can be approximated as a Markovian
two-state process. Suppose the system is sliding at time $t$. We call $p_s$ the
probability that it is still sliding at time $t+\Delta t$. Similarly, we call $q_s$
the probability of sliding at time $t+\Delta t$ knowing that the system is in a
jamming configuration at time $t$. Obviously, $p_s$ depends on $\alpha$ and $q_s$ on
$p$ and $\alpha$. For example, one has $q_s(p=0)=0$ and $q_s(p=1)=p_s$. 
\begin{figure}[htb]
\begin{center}
\epsfysize=7cm
\epsfbox{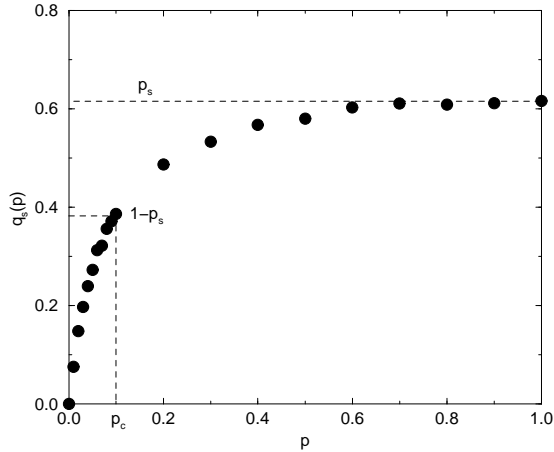}
\caption{\small The probability $q_s$ that a system in a jamming configuration at 
time $t$ becomes sliding at time $t+\Delta t$, depends on $p$. For fixed $R_c$, $b$
and $\alpha$, we call $p_c$ the critical value of $p$ such that $\alpha=\alpha_c$.
Three points of the curve $q_s(p)$ are analytically known: $q_s(0)=0$, $q_s(1)=p_s$
and $q_s(p_c)=1-p_s$. The whole curve has been obtained numerically for $R_c=0.5$,
$\alpha=\alpha_c(p=0.1)=0.914$ and $b=1$. \label{qgdep}}
\end{center}
\end{figure}
Figure \ref{qgdep} shows $q_s(p)$, as determined numerically. This simple two-state
model can be explicitly solved. The critical point is found to be when the
probability of sliding after jamming is equal to the probability of jamming after
sliding, i.e. $q_s(p) = 1-p_s$. At this point, the probability that $F$ increases is
equal to the probability that $F$ decreases, which implies indeed that $F$ behaves
as a random walk. Provided that the functions $p_s$ and $q_s$ are regular near the
critical line, one also finds the observed linear behaviour of $s$ and $1/\la \tau
\ra$. The fact that the temporal correlations are found to be small however means that our model fails to capture `precursor' effects before the slip, which have been observed experimentally in \cite{Gollub}. This is related to 
our simple rule where each grain can `move' under the influence of the external noise with equal probability. Finally, we also looked at 
the two following quantities: the distribution of
the heights of the slips, and the distribution of the intervals of time between two
slips. The tails of these distributions are found to be exponentially decaying.

In conclusion, we discussed how the Scalar Arching Model can 
be modified
to describe granular systems undergoing intermittent stick-slip motion. We showed
that such a system can present two different phases: a `slipping' stick-slip (or
sliding) phase, where the external pushing force remains finite, and a `sticking'
stick-slip (or jammed) phase. The transition between these two regimes is found to be of mean-field type.

\vskip 0.5cm 
Acknowledgments.
\vskip 0.5cm

We are grateful to E. Kolb ,T. Mazozi, J. Duran and E. Cl\'ement for having initiated
this work and for many stimulating discussions.

\end{document}